\documentclass{article}
\usepackage[english]{babel}
\usepackage{amsmath,amssymb,bbm}

\usepackage{tabularx}

\newcommand{\Iota}{\mathrm{I}}

\newcommand{\cdummy}{\cdot}

\newcommand{\tmop}[1]{\ensuremath{\operatorname{#1}}}
\newcommand{\tmrsub}[1]{\ensuremath{_{\textrm{#1}}}}

\newcommand{\tmscript}[1]{\text{\scriptsize{$#1$}}}

\newtheorem{theorem}{Theorem}

\newcommand{\seq}[2]{#1\tmrsub{1},#1\tmrsub{2}{\ldots},#1\tmrsub{#2}}

\usepackage{times}
\usepackage{graphicx} %
\usepackage{subfigure} 

\usepackage{natbib}

\usepackage{algorithm}
\usepackage{algorithmic}

\usepackage{hyperref}

\usepackage[accepted]{icml2014}

\usepackage{compactbib}
\usepackage[compact]{titlesec}

\usepackage{paralist}

\usepackage{enumitem}
\setlist[itemize,1]{leftmargin=0.5em,noitemsep,topsep=0pt}

\icmltitlerunning{Secure Friend Discovery via Privacy-Preserving and Decentralized Community Detection}

\begin{document} 

\twocolumn[
\icmltitle{Secure Friend Discovery via \\ Privacy-Preserving and Decentralized Community Detection}

\icmlauthor{Pili Hu}{hupili@ie.cuhk.edu.hk}
\icmlauthor{Sherman S.M. Chow}{smchow@ie.cuhk.edu.hk}
\icmlauthor{Wing Cheong Lau}{wclau@ie.cuhk.edu.hk}
\icmladdress{Department of Information Engineering,
            The Chinese University of Hong Kong}

\icmlkeywords{Privacy, Community Detection}

\vskip 0.3in
]

\begin{abstract}
  The problem of secure friend discovery on a social network has long been
  proposed and studied. The requirement is that a pair of nodes can make
  befriending decisions with minimum information exposed to the other party.
  In this paper, we propose to use community detection to tackle the problem
  of secure friend discovery.
  We formulate the first privacy-preserving and
  decentralized community detection problem as a multi-objective optimization.
  We design the first protocol to solve this problem, which transforms
  community detection to a series of Private Set Intersection (PSI) instances using
  Truncated Random Walk (TRW).
  Preliminary theoretical results show that our
  protocol can uncover communities with overwhelming probability and preserve privacy.
  We also discuss future works, potential extensions and variations.
\end{abstract}

\section{Introduction}

One important function provided by
social network is friend discovery.
The problem of finding people of the same attribute/ interest/ community has long been studied
in the context of social network.
For example, profile-based friend
discovery can recommend people who have similar attributes/ interests;
topology-based friend discovery can recommend people from the same community.

One special requirement of algorithms operating on social network is that it
must be privacy-preserving.
For example, social network nodes may be
willing to share their attributes/ interests with people having similar
profile; Or they may be willing to share their raw connections with people in
the same community. However, it is unfavourable to leak those private data to
arbitrary strangers. Towards this end, the friend discovery routine should only
expose minimal necessary information to involved parties.

In the current model of large-scale OSNs, service providers like Facebook
play a role of Trusted-Third-Party (TTP). The friend
discovery is accomplished as follows: 1) Every node (user) give his/her
profile and friend list to TTP; 2) TTP runs any sophisticated social network
mining algorithm (e.g. link prediction, community detection)
and returns the friend recommendations to only related users.
The mining
algorithm can be a complex one involving node-level attributes, netweork
topology, or both. Since TTP has all the data, the result can be very
accurate. This model is commercially viable and successfully deployed in
large-scale. However, recent arise of privacy concern motivates both
researchers and developers to pursue other solutions. Decentralized Social
Network (DSN) like Diaspora\footnote{\url{https://joindiaspora.com}}
has recently been proposed and implemented. Since
it is very difficult to design, implement and deploy a DSN
{\cite{datta2010decentralized}}, much research attention was focused on system
issues. We envision that the DSN movement will gradually grow with user's
increasing awareness of privacy. In fact, Diaspora,
the largest DSN up-to-date, has already
accumulated 1 million users. With the decentralized infrastructure
established, next question is: can we support accurate friend discovery under
the constraint that each node only observes partial information of the whole
social network? Note that the whole motivation of DSN is that single service
provider can not be fully trusted, so the TTP approach can not be re-used.
Towards this end, the computation procedure must be decentralized.

One common approach in literature to achieve decentralized and privacy-preserving friend
discovery is to transform it into a set matching problem. For the first type,
it is natural to represent one's attributes/ interests/ social activities in
form of a set
{\cite{zhang2012privatematching}}.
For the second type, one straightforward way is to represent one's friend
(neighbour) list in form of a set
{\cite{nagy2013knows}}.
In this way, both profile matching and common friend detection become a set
intersection problem. There exists one useful crypto primitive called Private
Set Intersection (PSI). Briefly and roughly speaking, given two sets
$\mathcal{W}_{1}$ and $\mathcal{W}_{2}$ held by two node $v_{1}$ and $v_{2}$,
PSI protocol can compute $| \mathcal{W}_{1} \cap \mathcal{W}_{2} |$ without
letting either $v_{1}$ or $v_{2}$ know other party's raw input. Resaerchers
have proposed PSI schemes based on commutative encryption
{\cite{agrawal2003information}}, oblivious
polynomial evaluation
{\cite{freedman2004efficient}}
oblivious psudorandom function {\cite{freedman2005keyword}},
index-hiding message encoding {\cite{manulis2010privacy}}, hardware
{\cite{hazay2008constructions}}
or generic construction {\cite{huang2012private}}
using garbled circuit {\cite{yao1982protocols}}.
The
aforementioned privacy-preserving profile matching/ common friend detection
protocols are variants of PSI protocols in terms of output, adversary model,
security requirement and efficiency.

One major drawback of all the above works is that they can not fully utilize
the topology of a social network.
Firstly, profile is just node-level
information and not always available on every social network. 
On the contrary, topology (connections/
friendship relations) is the fundamental data available on social networks.
Secondly, common friend is just one topology-based approach 
and it only works for nodes within 2-hops. 
In fact, our previous investigation
showed that common friend heuristic has a moderate precision and low recall for
discovering community-based friendship {\cite{hu2013cd2hop}}. 
This result is
unsurprising because %
a community can easily span multiple hops. 
Towards this end, 
we focus on extending traditional secure friend discovery beyond 2-hops via community detection. 
Note that topology-only community detection
{\cite{clauset2004finding}}
{\cite{blondel2008fast}}
{\cite{raghavan2007near}}
{\cite{leung2009towards}}
{\cite{agarwal2008modularity}}
{\cite{coscia2012demon}}
{\cite{soundarajanuse2013local}}
is a classical problem under centralized and non privacy-preserving setting,
i.e. a single-party possesses the complete social graph and does arbitrary computation.
Although one can translate those algorithms into a
privacy-preserving and decentralized protocol using generic garbled circuit
construction {\cite{yao1982protocols}}, the computation and communication cost
renders it impractical in the real world.
To design an efficient scheme, we need
to consider community detection accuracy and privacy preservation as a whole.
A tradeoff among accuracy, privacy and efficiency can also be made when necessary.

To summarize, this paper made the following contributions:
\begin{itemize}
    \item We proposed and formulated 
        the first \emph{privacy-preserving} and \emph{decentralized} community
  detection problem, which largely improves the recall of topology-based
  friend discovery on Decentralized Social Networks.

  \item We designed the first protocol to solve this problem.
      The protocol transforms the community detection
      problem to a series of Private Set Intersection (PSI) instances
    via Truncated Random Walk (TRW).
    Preliminary results show that the protocol can
  uncover communities with overwhelming probability and preserve privacy.

  \item We propose open problems and discuss future works, extensions and variations in the end.
\end{itemize}

\section{Related Work}

First type of related work is Private Set Intersection (PSI) as they are
already widely used for secure friend discovery.
Second type of related work
is topology-based graph mining.
Although our problem is termed ``community detection'',
the most closely related works are actually topology-based Sybil defense.
This is because previous community detection problems are mainly considered
under the centralized scenario. On the contrary, Sybil defense scheme sees
wide application in P2P system, so one of the root concern is decentralized
execution.
Note, there exist some distributed community detection works 
but they can not be
directly used because nodes exchange too much information. For example
{\cite{hui2007distributed}} allow nodes to exchange adjacency lists and
intermediate community detection results, which directly breaks the privacy
constraint that we will formulate in following sections.
Due to space limit, a detailed survey of related work is omitted.
Interested readers can see
community detection surveys {\cite{fortunato2010community}}{\cite{xie2011overlapping}}
and Sybil detection surveys {\cite{yu2011sybiltut}}{\cite{alvisi2013sok}}.

\section{Problem Formulation}

The notion of community is that intra-community is dense
and inter-community linkage is sparse.
In this section, we first review classical community detection formulations under
centralized scenario and our previous formulation under decentralized
scenario. Then we formulate the privacy-preserving version. To make the
problem amenable to theoretical analysis, we consider a Community-Based Random
Graph (CBRG) model in the last part. 

\subsection{Previous Community Detection Formulations}

Classical community detection is formulated as a clustering problem. That is,
given the full graph $G= (V,E)$, partition the vertex set into $K$ subsets
$\seq{S}{K}$ (a partitioning),
such that $\cap_{i=1}^{K} S_{i} = \emptyset$ and $\cup_{i=1}^{K}
S_{i} =V$. A quality metric $Q ( \{ S_{1} , \ldots ,S_{K} \} )$ is defined
over the partitions and a community detection algorithm will try to find a
partitioning that maximize or minimize $Q$ depending on its nature. This is for
non-overlapping community detection and one can simply remove the constraint
$\cap_{i=1}^{K} S_{i} = \emptyset$ to get the overlapping version.
Note that $Q$ is only an artificial surrogate to the axiomatic notion of
community. The maximum $Q$ does not necessarily corresponds to the best
community. 
However, the community detection problem becomes tractable via
well-studied optimization frameworks by assuming a form of $Q$
e.g. Modularity, Conductance.
Most classical works are along this line mainly due to the lack of ground-truth
data at early years.

Now consider the decentralized scenario. One node (observer) is limited to its
local view of the whole graph. It is unreasonable to ask for a global
partitioning in terms of sets of nodes. 
The tractable question to ask is:
whether one node is in the same community as the observer or not? This gives a
binary classification formulation of community detection {\cite{hu2013cd2hop}}.
The result of community detection with respect to a
single observer can be represented as a length-$| V |$ vector.
Stacking all those vectors together, we can get a community encoding matrix
{\cite{zhong2014pbcm}}:
\[ M_{i,j} = \left\{ \begin{array}{ll}
     1 & \exists S_{k} ,s.t. v_{i} \in S_{k} ,v_{j} \in S_{k}\\
     0 & \tmop{else}
   \end{array} \right. \label{eq:model-community-encoding-matrix} \]
This matrix representation is subsumed by partitioning representation in general case.
If restricted to non-overlapping case, the two representations are equivalent. 
Since $M$ encodes all pair-wise outcome, it is immediately
useful for friend discovery application. In what follows, we will define
accuracy and privacy in terms of how well $M$ can be learned by nodes or
adversary.

\subsection{Privacy-Preserving Community Detection}

In this initial study, we focus on non collusive passive adversary.
That is,
DSN nodes all execute our protocol faithfully but they are curious to infer
further information from observed protocol sequence.
We use a single non-collusive sniff-only adversary to capture this notion.
The system components are as follows:
\begin{itemize}
  \item Graph: $G= ( V,E )$. The connection matrix is denoted as $C$, where
  $C_{i,j} =1$ if $( v_{i} ,v_{j} ) \in E$; otherwise, $C_{i,j} =0$. The
  ground-truth community encoding matrix is denoted as $M^{g}$, which is
  unknown to all parties at the beginning. For simplicity of discussion, we
  assume the nodes identifiers, i.e. $V$, is public information.
  
  \item Nodes: $v_{1} , \ldots ,v_{| V |} \in V$. A node's initial knowledge
  is its own direct connections, i.e. $N ( v_{i} ) = \{ v_{j} | ( v_{j} ,v_{i}
  ) \in E \}$. 
  Nodes are fully honest. 
  Their objective is to maximize the accuracy of detecting $M$. 
  Eventually, a node $v_{i}$ can get full row
  (column) in $M$ denoted by $M_{i,:}$ ($M_{:,i}$).
  Depending on the protocol choice,
  relevant cells in $M$ can be made available immediately or on-demand.
  
  \item Adversary: $A$. It can passively sniff on one node $v_{a} \in V$. $A$
  will observe all protocol sequence related with $a$, including initial
  knowledge $N ( v_{a} )$ and the community detection result $M_{a,:}$. $A$'s
  objective is to maximize successful rate in guessing $M^{g}$ and $C$, using
  any Probabilistic Polynomial Algorithms (PPA).
  Note, the full separation of Nodes and Adversary is for ease of discussion.
  In real DSN, this passive attacker can be a curious user
  who wants to infer more information of the network.
\end{itemize}
As protocol designer, our objectives are:
\begin{itemize}
  \item Accurately detect community after execution of the
  protocol, i.e. making $M$ and $M^{g}$ as close as possible.
  
  \item Limit the successful rate of adversary's guessing of $M^{g}$ and $C$,
  under the condition that $A$ gets the protocol sequence on node $v_{a}$
  and makes best guess via PPA.
\end{itemize}
One can see that our problem is multi-objective in nature. 
The accuracy part is a maximization problem
and the privacy part is a is min-max problem.
Formal definition is given in Eq.~\ref{eq:model-pcd-multi-objective}.

\begin{figure*}[t!]
\begin{equation}
  \max_{\tmscript{\begin{array}{l}
    \tmop{Find}   \tmop{Protocol} ,\\
    M= \tmop{Protocol} ( G )
  \end{array}}} \left( \begin{array}{l}
    \tmop{Succ} ( M,M^{g} ,V \times V ) ,\\
    - \left( \max_{\tmscript{\begin{array}{l}
      \tmop{Algo} \in \tmop{PPA} ,\\
      a \xleftarrow{\$} V,\\
      C^{A} ,M^{A} \leftarrow \tmop{Algo} ( \tmop{Protocol} ,I_{a} )
    \end{array}}} \left( \begin{array}{l}
      \tmop{Succ} ( C^{A} ,C,R_{a}^{C} ) ,\\
      \tmop{Succ} ( M^{A} ,M^{g} ,R_{a}^{M} )
    \end{array} \right) \right)
  \end{array} \right) \label{eq:model-pcd-multi-objective}
\end{equation}
\vspace{-2em}
\end{figure*}

In this formulation, ``Protocol'' is an abstract notation of the protocol
specification, not protocol execution sequence. $I_{a}$ is the information
observed by adversary, which is dependent on Protocol. $\tmop{Succ} ( B^{1}
,B^{2} ,R )$ is the measure of successful rate with symbols defined as
follows:
\begin{itemize}
  \item $B^{1} ,B^{2} \in \{ 0,1 \}^{| V | \times | V |}$ are two $\{ 0,1 \}$
  matrix in the same size as $M$ and $C$.
  
  \item $R \subseteq V \times V$ is the challenge relations.
  
  \item To measure how close are the two matrix over the challenge set, we
  use the successful rate:
  \[ \tmop{Succ} ( B^{1} ,B^{2} ,R ) = \Pr \left\{ B^{1}_{i,j} =B^{2}_{i,j} |
     ( v_{i} ,v_{j} ) \xleftarrow{\$} R \right\} \]
  That is, how likely a randomly selected pair of nodes from $R$ will have the
  same value in $B^{1}$ and $B^{2}$.
\end{itemize}
For the accuracy part, we define the challenge relation as $V \times V$
because we want the result to be accurate for all nodes.
For the privacy part,
we define the challenge relation as $R_{a}^{C} =R_{a}^{M} =(V-U ( a ))\times(V-U(a))$,
where $U ( a )$ denotes the set of nodes in the same community as $a$. 
The reason to exclude nodes from the same community is obvious.
Since adversary will
get $M_{a} ,:$ after protocol execution, it already knows the community
membership of $U ( a )$. Given the knowledge of community, one can make more
intelligent guess of the connections. This is made clear in later discussions.

\subsection{Community-Based Random Graph (CBRG) Generation Model}

Before proceed, we remark that the problem defined in Eq.
\ref{eq:model-pcd-multi-objective} is hard even without the privacy-preserving objective.
In other words, the community detection problem (accuracy)
has not been fully solved even under the TTP scenario.
To improve the accuracy, researchers have already used heavy mathematical programming tools, try to
incorporate more side information, develop problem-specific heuristics, or
perform heavy-duty parameter tuning.
To make our problem amenable to theoretical analysis, we consider a
Community-Based Random Graph (CBRG) model in this paper.
Let $M^{g}$ be
the ground-truth community encoding matrix. We generate the random connection
matrix as follows: 1) $\Pr \{ C_{i,j} =1 \} =p$ if $M^{g}_{i,j} =1$ ($v_{i}$
and $v_{j}$ are in the same community); $\Pr \{ C_{i,j} =1 \} =q$ otherwise.
There are $K$ communities and
each of size $c$, so the total number of vertices is $| V | =Kc$. We denote
such a random graph as $\tmop{CBRG} ( K,c,p,q )$. 
One example ground-truth community encoding matrix and the expected connection matrix are illustrated
in Fig.~\ref{fig:cbrg-illustration}.

\begin{figure}
  \[ M^{g} = \left[ \begin{array}{llll}
        1 &  &  &   \\
        1 & 1 &  &   \\
        0 & 0 & 1 &   \\
        0 & 0 & 1 & 1 
     \end{array} \right] ,\mathbbm{E} [ C ] = \left[ \begin{array}{llll}
        p &  &  &   \\
        p & p &  &   \\
        q & q & p &   \\
        q & q & p & p 
     \end{array} \right] \]
\vspace{-2em}
  \caption{Illustration of community-based random graph generation. $K=2,c=2$}
  \label{fig:cbrg-illustration}
\vspace{-1.5em}
\end{figure}

\section{Proposed Scheme}

In this section, we present our protocol and main results.

\subsection{Protocol Design}

Our protocol involves the two stages:

\begin{itemize}
  \item Pre-processing is done via Truncated Random Walk.
  Every node send out $W$ random walkers,
  $w_{1}^{v_{i}} , \ldots ,w_{W}^{v_{i}}$, with time-to-live (TTL) values
  $l_{1}^{v_{i}} , \ldots ,l_{W}^{v_{i}}$ initially set to $L$. 
  Upon receiving a Random Walker (RW) $w$, 
  the node records the ID of $w$, 
  deducts its
  TTL $l$, and sends it to a random neighbour if $l>0$.
  At the end of this
  stage, each node $v_{i}$ accumulated a set of random walker IDs
  $\mathcal{W}_{i}$. 
  With proper parameters $W$ and $L$, the truncated random
  walker issued by $v_{i}$ will more likely reach other nodes in the same community as $v_{i}$. 
  So by inspecting the intersection size of
  $\mathcal{W}_{i}$ and $\mathcal{W}_{j}$,
  we can answer whether $v_{i}$ and $v_{j}$ are in the same community.
  This essentially
  transforms the community detection problem to a set intersection problem.
  
  \item To uncover the relevant cells in pairwise community encoding matrix
      $M$, we only need to perform Privacy Set Intersection (PSI) on two sets.
      PSI schemes differ in their flavours:
      1) reveal intersection set (PSI-Set);
      2) reveal intersection size (PSI-Cardinality);
      3) reveal whether intersection size is greater than a threshold (PSI-Threshold).
      We use the 3rd type PSI in our construction,
      which can be implemented by adapting \cite{zhang2012privatematching}.
      In what follows, we just assume existence of such a crypto primitive:
      it computes $\Iota [ | \mathcal{W}_{i} \cap \mathcal{W}_{j} | \geqslant T ]$
      without leaking extra information.
\end{itemize}
One can see that the scheme is decentralized by design. We only need to argue
its community detection accuracy and the privacy-preserving property.

\subsection{Summary of Theoretical Guarantees}

The intuition of our proof is as follows:
\begin{itemize}
  \item Truncated Random Walk will be mostly limited to one community, if the
  axiomatic notion of ``community'' holds. More precisely, as long as $p$ is
  enough larger than $\beta_{1} = ( K-1 ) q$, there will be enough
  difference in intersection size for nodes coming from the same and different
  communities. In this case, we can set proper threshold to ensure low error
  rate.
  
  \item Observe two facts about privacy objective: 1) most protocol sequence
  the adversary observed comes from its own community; 2) we exclude $A$'s
  community from challenge relations. In order to make better-than-priori
  guesses, $A$ at least need to observe some other nodes from protocol sequence. The
  number of nodes from $V-U ( a )$ can be observed is limited. Even if we
  assume adversary can make good use of the information (captured by
  coefficient $\gamma_{M} , \gamma_{C} \in [ 0,1 ]$), this small advantage is
  averaged out over a large challenge relation set.
\end{itemize}
The detailed proof is omitted and the main results are summarized in the
following theorem.

\begin{theorem}
  Our protocol guarantees:
  \begin{itemize}
    \item False Positive Rate: $$\Pr \{ | \mathcal{W}_{i} \cap \mathcal{W}_{j}
    | \geqslant T_{1} |M^{g}_{i,j} =0 \} \leqslant \frac{\phi WL ( L+1
    )^{2}}{2 ( K-1 ) T_{1}}$$
    
    \item False Negative Rate: ($\mu =cWP$)
        $$\Pr \{ | \mathcal{W}_{i} \cap \mathcal{W}_{j}
    | \leqslant T_{2} |M^{g}_{i,j} =1 \} \leqslant e^{- \mu ( 1-T_{2} / \mu
    )^{2} /2}$$
    
    \item Adversary's advantage:
    \begin{eqnarray*}
      \tmop{Adv} ( M^{A} ,M^{g} ,R^{M}_{a} ) & \leqslant & \gamma_{M}
      \frac{4W ( L+1 )}{( K-1 ) c} \\
      \tmop{Adv} ( C^{A} ,C^{g} ,R^{C}_{a} ) & \leqslant & \gamma_{C}
      \frac{4W ( L+1 )}{( K-1 ) c} 
    \end{eqnarray*}
  \end{itemize}
\end{theorem}

In the theorem, $\tmop{Adv} ( B^{1} ,B^{2} ,R ) = \tmop{Succ} ( B^{1} ,B^{2}
,R ) - \tmop{Prior} ( B^{1} ,B^{2} ,R )$. $\tmop{Prior} ( B^{1} ,B^{2} ,R )$
denotes the probability to make successful guess based on mere prior
information of $B^{2}$. For example, suppose $B^{2}$ contains $1$ as
majority, i.e. $\Pr \left\{ B^{2}_{i,j} =1|i,j \xleftarrow{\$} R \right\}
=P>0.5$. The best guess is to let $B^{1}_{i,j} =1, \forall i,j \in R$. One can
show that the success probability is $P$ and this strategy is optimal if no
other information is available. Due to the specifics of our problem, adversary
can make more intelligent guesses than random $\{ 0,1 \}$ bit. Towards this
end, the advantage is defined with respsect to successful rate of this
priori-based strategy.

\subsection{One Instantiation}

Due to the specifics of the problem,
both accuracy and privacy guarantees are parameterized.
To give an intuitive view of what can be achieved,
consider one instantiation of CBRG: $K=100$ (\# of
communities), $c=500$ (\# of nodes in one community), $p=0.5$ (intra-community
edge generation probability), $\beta_{1} =q ( K-1 ) =0.05$, $q=0.0005$
(inter-community edge generation probability).

We can set protocol parameters as follows: $W=100$ (\# of RWs
issued by one node), $L=3$ (length of RW) and $T=61$ (threshold of intersection size).
This gives us following accuracy and privacy guarantees:
\begin{itemize}
  \item False Negative Rate: $\leqslant 1.9 \times 10^{-22}$
  
  \item False Positive Rate: $\leqslant 0.066$
  
  \item Advantage for guessing $M$: $\leqslant 0.032 \times \gamma_{M}$
  
  \item Advantage for guessing $C$: $\leqslant 0.032 \times \gamma_{C}$
\end{itemize}
One can see that our proposed protocol can accurately detect community and
preserve privacy given proper parameters. 
Note first that above $W$ and $L$
are casually selected by heuristics, which have not been jointly optimized. Note second
that the FPR and FNR can be exponentially reduced by repeated experiments,
which only maps to a linear increase in $W$. The example in this section is
only to demonstrate the effectiveness of our protocol and a full exploration
of design space is left for future work.

\section{Conclusion, Discussion and Future Work}

We formulated the privacy-preserving community detection problem in this paper
as a multi-objective optimization. We proposed a protocol based on Truncated
Random Walk (TRW) and Private Set Intersection (PSI). We have proven that our
protocol detects community with overwhelming probability and preserves privacy.
Exploration of the design space and thorough
experimentation on synthesized/ real graphs are left for future work.
In following parts of this early report,
we discuss several simpler candidate protocols and how they fail to meet our objective.
This help to demonstrate the rationale of our formulation and protocol design.

\subsection{Simpler But Weaker Protocols}

Suppose we change the protocol such that 
$v_{i}$ and $v_{j}$ first exchange $\mathcal{W}_{i}$ and
$\mathcal{W}_{j}$ and then run any intersection algorithm separately. 
After uncovering all related cells in $M$, 
adversary knows $\mathcal{W}_{i} , \forall i=1, \ldots , |
V |$. $A$ can directly calculate $| \mathcal{W}_{i} \cap \mathcal{W}_{j} | ,
\forall i,j$. This allows adversary to guess $M$ perfectly.
From the community membership, $A$ can further
infer links because intra-community edge generation probability and
inter-community generation probability are different. This already allows
better guess than using global prior of $C$. Furthermore, inferring links from
measurements is a classical well-studied topic called Network Tomography. $A$
can actually re-organize $\mathcal{W}_{i}$'s into a list of size-$L$ sets,
each representing the nodes traversed by a RW. 
Researchers have shown
that links can be inferred from this co-occurrence data with good
accuracy, e.g. NICO {\cite{rabbat2008network-nico}}.

Another natural thought to protect non-common set elements is via hashing. Suppose
there exists a cryptographic hash $h ( \cdummy )$. We define $\mathcal{H}_{i}
= \{ h ( w ) |w \in \mathcal{W}_{i} \}$. Now, two nodes just compare
$\mathcal{H}_{i}$ and $\mathcal{H}_{j}$ in the community uncover stage. This
can protect true identities of the RWs if their ID space is large enough.
However, it does not prevent adversary from intelligent guess of $M$ and $C$.
Methods noted in previous paragraph can also be used in this case.

In our protocol, we used the PSI-Threshold version. That is, given
$\mathcal{W}_{i}$ and $\mathcal{W}_{j}$, the two parties know nothing except
for the indicator $\Iota [ | \mathcal{W}_{i} \cap \mathcal{W}_{j} | \geqslant
T ]$. Two weaker and widely studied variations are: PSI-Cardinality and PSI-Set.
Consider PSI-Set. The adversary now only knows elements in the
intersection. Based on his own $\mathcal{W}_{a}$ and PSI-Set protocol
sequence, he can get $\mathcal{W}_{i} \cap \mathcal{W}_{j} \cap
\mathcal{W}_{a} , \forall i,j$. $A$ can calculate the
probability that a RW $w$ tranverses both $v_{i}$ and $v_{j}$ conditioned on
$w$ tranverses $v_{a}$. Based on this information, $A$ can adjust threshold
$T_{1}$ and $T_{2}$ to accurately detect communities. The derivation is
similar to our protocol in this paper but more technically involved, which is also
left as future work. The bottom line is that PSI-Set leaks enough information
for more intelligent guesses.
As for PSI-Cardinality, we are not
sure at present what an adversary can do with
$| \mathcal{W}_{i} \cap \mathcal{W}_{a} |, \forall i$.
Since the two variants leak more information and might be potentially exploited, 
we use PSI-Threshold in our protocol.

\subsection{Open Problems}

Following are some open problems of privacy-preserving community detection:
\begin{itemize}
  \item If we allow a small fraction of nodes to collude, how to define a
  reasonable security game? What privacy-preserving result can we achieve?
  
  \item Current scheme requires all nodes to re-run the protocol, if there is
  any change in the topology, e.g. new node joins or new friendship
  (connection) is formed. Is it possible to find a privacy-preserving
  community detection scheme that can be incrementally updated?
  
  \item The privacy preservation of our proposed protocol is dependent on
  graph size. One root cause is that we only leveraged crypto primitives in
  the Private Set Intersection (PSI) part. The simulation of Truncated Random
  Walk (TRW) is done in a normal way. Since random walk is a basic construct
  in many graph algorithms, it is of interest know how (whether or not) nodes
  can simulate Random Walk in a decentralized and privacy preserving fashion.

\end{itemize}

\end{document}